\documentclass{moriond}

\usepackage{amsmath}

\def\Journal#1#2#3#4{{#1} {\bf #2}, #3 (#4)}

\def\PRD{{\em Phys. Rev.} D}

\def\be{\begin{equation}}
\def\ee{\end{equation}}
\def\bea{\begin{eqnarray}}
\def\eea{\end{eqnarray}}

\def\k{{\bf k}}
\def\z{{\bf z}}
\def\q{{\bf q}}

\newcommand{\Photo}{}

\begin{document}
\vspace*{4cm}
\title{Precision cosmology from large-scale structure of the Universe}

\author{A. Chudaykin$^{1,2}$}

\address{$^{1}$Department of Physics \& Astronomy, McMaster University,\\ 
	1280 Main Street West, Hamilton, ON L8S 4M1, Canada}
\address{$^{2}$Institute for Nuclear Research of the
	Russian Academy of Sciences, \\ 
	\normalsize \it  60th October Anniversary Prospect, 7a, 117312
	Moscow, Russia}

\maketitle\abstracts{
Large scale structure of the Universe 
becomes a leading source
of precision cosmological information. 
We present two particular tools
that can be used in cosmological 
analyses of the redshift space 
galaxy clustering data: 
a new open-source code \texttt{CLASS-PT}
and the theoretical error approach. 
\texttt{CLASS-PT} computes one-loop power auto- and cross-power spectra for matter fields and biased tracers in real and redshift spaces. 
We show that the code meets the 
precision standards set by the 
upcoming high-precision large-scale
structure surveys. 
The theoretical error likelihood approach  allows one to analyze galaxy clustering data without having to measure 
the scale cut $k_{\rm max}$.
This approach takes into account 
that theoretical uncertainties 
affect parameter estimation gradually, 
which helps optimize
data 
analysis and ensures that 
all available cosmological information
is extracted. 
}

\section{Introduction}
\label{sec:intro}

Large-scale structure (LSS) data becomes an increasingly important source of cosmological information. 
Very soon it will become
competitive with
the cosmic microwave background (CMB)
data analysis~\cite{Chudaykin:2019ock}.
In contrast with the CMB, the LSS data analysis is complicated by non-linear clustering which is very hard to predict analytically. This problem 
is resolved in the effective
field theory of large-scale structure~\cite{Baumann:2010tm}, 
and its various extensions, such as time-sliced perturbation theory~\cite{Blas:2015qsi}, which 
allow one to systematically 
model various short-scale phenomena on mildly nonlinear scales.


In the first part of this note we present a new code \texttt{CLASS-PT} that embodies an end-to-end calculation
of one-loop power spectra for matter field and bias tracers using the state-of-the-art perturbation theory approach~\cite{Chudaykin}. Even though \texttt{CLASS-PT} is based on the well-known theoretical framework~\cite{Ivanov:2019pdj}, it brings several novelties. First, it uses a new FFTLog method to 
efficiently
compute convolution integrals~\cite{Simonovic:2017mhp}. Second, another crucial advantage of \texttt{CLASS-PT}
is an accurate and efficient 
description of the BAO wiggles using the infrared (IR) resummation of long-wavelength contributions~\cite{Blas:2016sfa,Ivanov:2018gjr}. This is particularly important for data analysis, 
where the BAO encapsulate a significant portion of cosmological information.

In the second part, we introduce a novel data analysis technique based on the theoretical error covariance~\cite{Chudaykin2}, which allows one to avoid uncertainties in the theoretical estimates of higher-order nonlinearities.
We demonstrate that this approach yields unbiased constraints on all cosmological parameters. In addition, we also show that the theoretical error effectively optimizes the choice of $k_{\max}$ in realistic data analyses.

\section{\texttt{CLASS-PT}}
\label{sec:class}

Let us start with theoretical modeling of galaxy power spectrum in redshift space. The radial position of the galaxies in real surveys are distorted by the peculiar velocity field. It introduces so-called redshift-space distortions (RSD). We will work in the plane-parallel approximation, where the mapping between the redshift and real spaces is entirely parameterized by the cosine of the angle between the line-of-sight $\hat\z$ and the wavevector of a given Fourier mode $\k$, $\mu\equiv (\hat\z\cdot \k)/k$. This setup allows one to express the one-loop redshift-space spectrum in the following simple form
\be
\label{eq:mastz}
\begin{split}
	P_{\rm gg,RSD}(z,k,\mu)= & Z^2_1(\k)
	P_{\text{lin}}
	(z,k)+ 2\int_{\q}Z^2_2(\q,\k-\q)
	P_{\text{lin}}(z,|\k-\q|)
	P_{\text{lin}}(z,q)\\
	& + 6Z_1(\k)P_{\text{lin}}(z,k)\int_{\q}Z_3(\q,-\q,\k)P_{\text{lin}}(z,q)\\
	& + P_{\text{ctr,RSD}}(z,k,\mu)
	+ P_{\epsilon\epsilon,\text{RSD}}(z,k,\mu)\,,
\end{split}
\ee
where the redshift-space kernels can be found in \cite{Ivanov:2019pdj}. $P_{\text{ctr,RSD}}(z,k,\mu)$ represents counterterm contributions in redshift space which in the leading order can be written as follows
\be
\label{eq:pctrrsd}
\begin{split}
	P_{{\rm ctr, RSD,}} (z,k,\mu) = & -2\tilde c_0(z) k^2 P_{\text{lin}}(z,k)  -2\tilde c_2(z) f(z) \mu^2 k^2 P_{\text{lin}}(z,k)\\
&	-2\tilde c_4(z) f^2(z)\mu^4 k^2 P_{\text{lin}}(z,k)-2b_4(b_1+f\mu^2)^2 k^4P_{\rm lin}(k) \,,
\end{split}
\ee
$P_{\epsilon\epsilon,\text{RSD}}(z,k,\mu)$ denotes the stochastic contribution which in the redshift space has the following structure 
\be
P_{\epsilon\epsilon,{\rm RSD}}(z,k,\mu)=P_{\rm shot}(z)+ a_0(z) k^2 + a_2(z) \mu^2 k^2 \,,
\ee
where $P_{\rm shot}(z)$ describes a constant shot noise and the additional terms represent scale-dependent contrinutions to the monopole and the quanrupole moments of power spectrum.

\begin{figure}[t!]
	\includegraphics[width=0.49\linewidth]{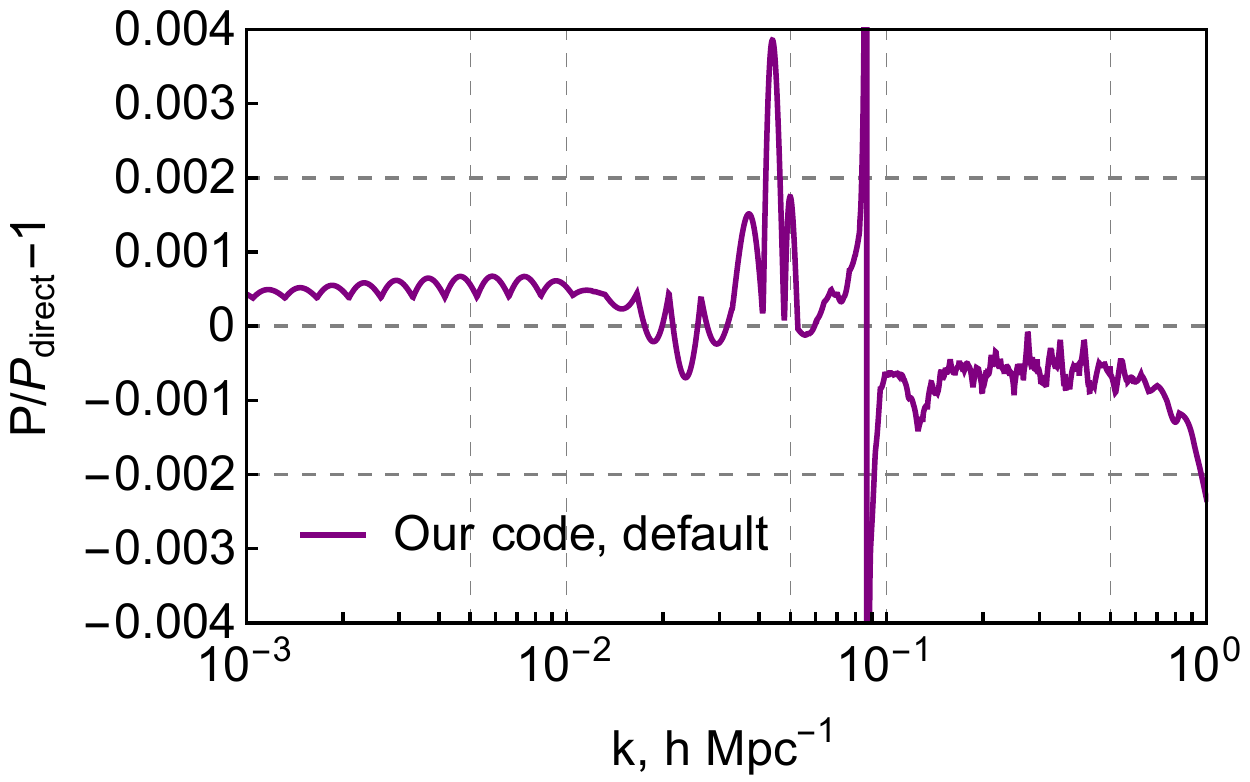}
	\includegraphics[width=0.48\linewidth]{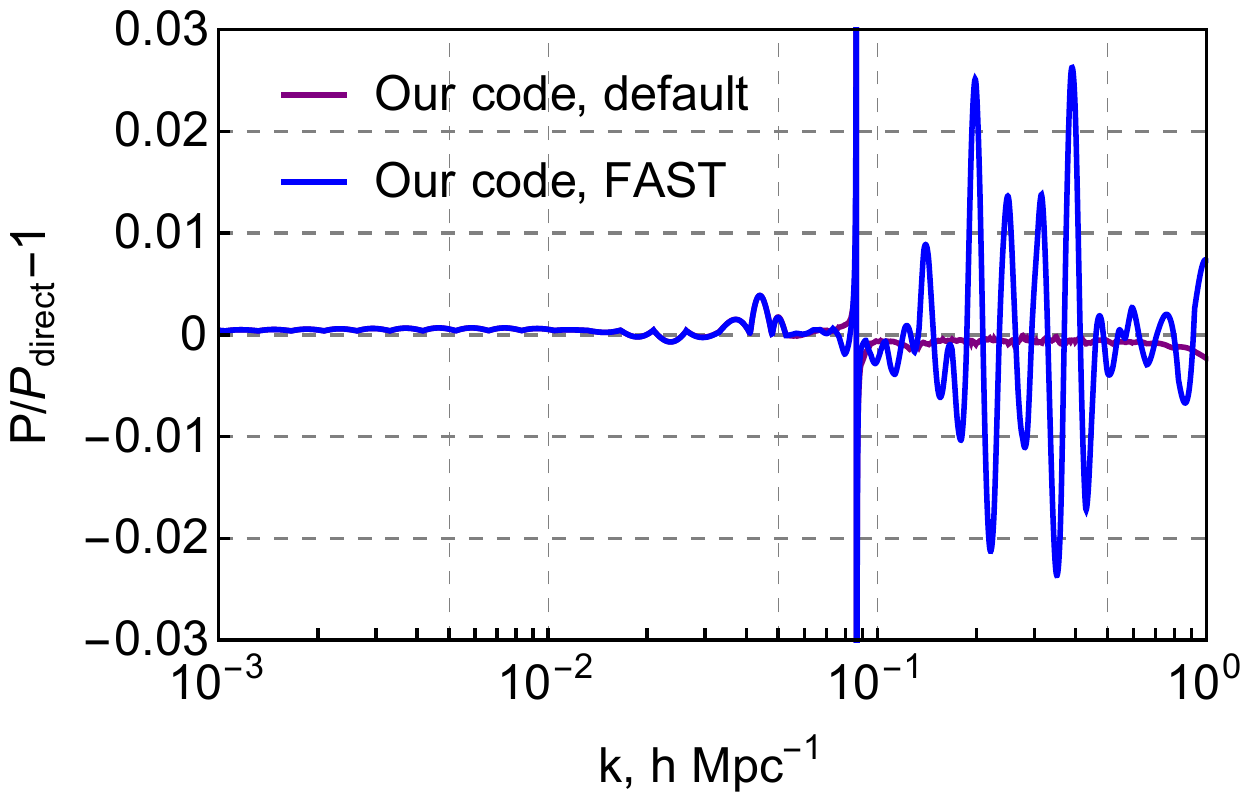}
	\caption[]{Residuals between the one-loop correction to the  matter power spectrum and its direct numerical evaluation for the default settings (left panel) and in the FAST mode (right panel).}
	\label{fig:1}
\end{figure}

The full angular dependence of the redshift-space power spectrum can be encoded in a number of multipoles using the following relation
\be
P_{\rm gg,RSD}(z,k,\mu) =\sum_{\ell \; {\rm even}} {\cal L}_\ell(\mu) P_{\ell}(z,k)\,,
\ee
Explicit expressions for $P_\ell$ can be found in \cite{Chudaykin}.
Let us briefly discuss the accuracy of calculations in \texttt{CLASS-PT}. In Fig. \ref{fig:1} we show the residuals between evaluation of the one-loop correction to the matter power spectrum with the FFTLog method and its direct numerical calculation for two different precision settings.
In the default regime we implement the FFTLog method for the grid with $N_{\rm FFTLog}=256$ harmonics. In the FAST mode we create a grid of lower dimension $N_{\rm FFTLog}=128$ that significantly speeds up our calculations. We found that the default choice of $N_{\rm FFTLog}=256$ provides $\sim0.1\%$ accuracy which is sufficient for future galaxy surveys. 
The FAST mode has somewhat worse accuracy
of the one-loop calculation, around $\sim1\%$. Still, it translates to the $\mathcal{O}(0.1\%)$ accuracy on the \textit{total} power spectrum. Therefore, the FAST mode is sufficient for the analysis of current LSS data.

Let us discuss the performance of \texttt{CLASS-PT}. Tab. \ref{table0} summarizes the run times for various tests in two regimes (default and FAST).
\begin{table*}[t!]
	\begin{tabular}{|c||c|c|c|c|c|} \hline
		Run  &  Real space & IR  resum. &  RSD  &  
		IR+RSD & IR+RSD+AP
		\\ [0.0cm]
		\hline 
		\multicolumn{6}{|c|}{Default mode}   \\
		\hline 

		\hline 
		Matter &  0.036 (0.036) & 
		0.175 (0.036)
		& 0.375 (0.375) 
		& 0.75 (0.62) & 0.76 (0.63) \\ \hline
		Tracers  & 0.21 (0.21) 
		& 0.35 (0.21)
		& 0.89 (0.89)  
		& 1.27 (1.12) &  1.30 (1.14)\\ 
		\hline 
		\multicolumn{6}{|c|}{FAST mode}   \\
		\hline 
		Matter &  $6.3~(6.1)\times  10^{-3}$ & 
		0.14 (0.0061)
		& 0.063 (0.061) 
		& 0.22 (0.09)  & 0.22 (0.09)  \\ \hline
		Tracers  & 0.033 (0.034) 
		& 0.17 (0.034)
		& 0.14 (0.14)  
		& 0.31 (0.18) &  0.31 (0.18)\\ 
		\hline 
		\hline 
	\end{tabular}
	\caption{Performance of the code in the baseline and FAST precision modes. 
		We show the execution time 
		in [sec.] as follows: $t_{\rm full}(t_{\rm FFTLog})$, where 
		$t_{\rm full}$ is the full evaluation time taken by the non-linear module,
		and $t_{\rm FFTLog}$ is the time elapsed during the matrix multiplication with FFTLog method.
	}
	\label{table0}
\end{table*}
Our results show that the galaxy power spectra in redshift space can be calculated over $1.3$ seconds for high precision settings. In the FAST mode the execution time reduces to $0.3$ seconds.

\begin{figure}
	\includegraphics[width=0.48\linewidth]{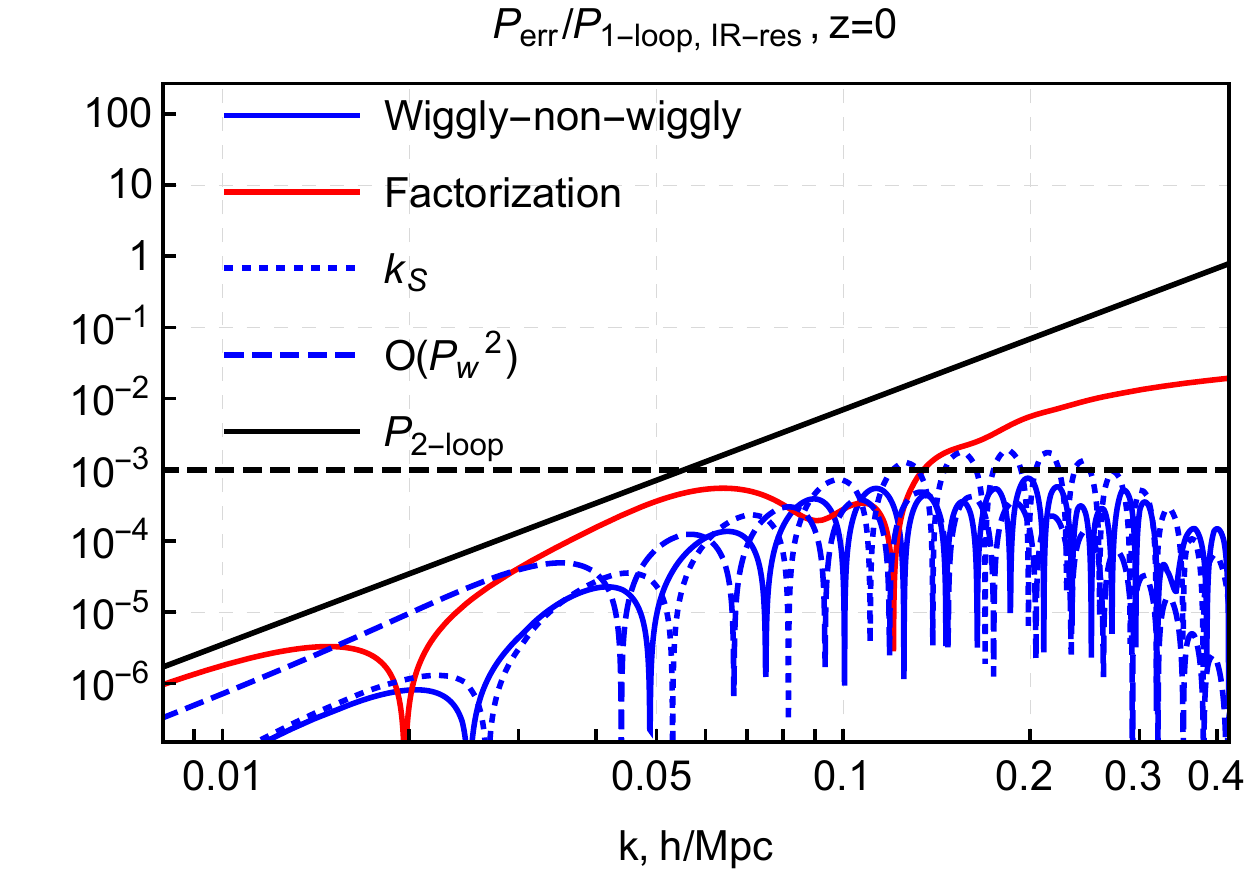}
	\includegraphics[width=0.48\linewidth]{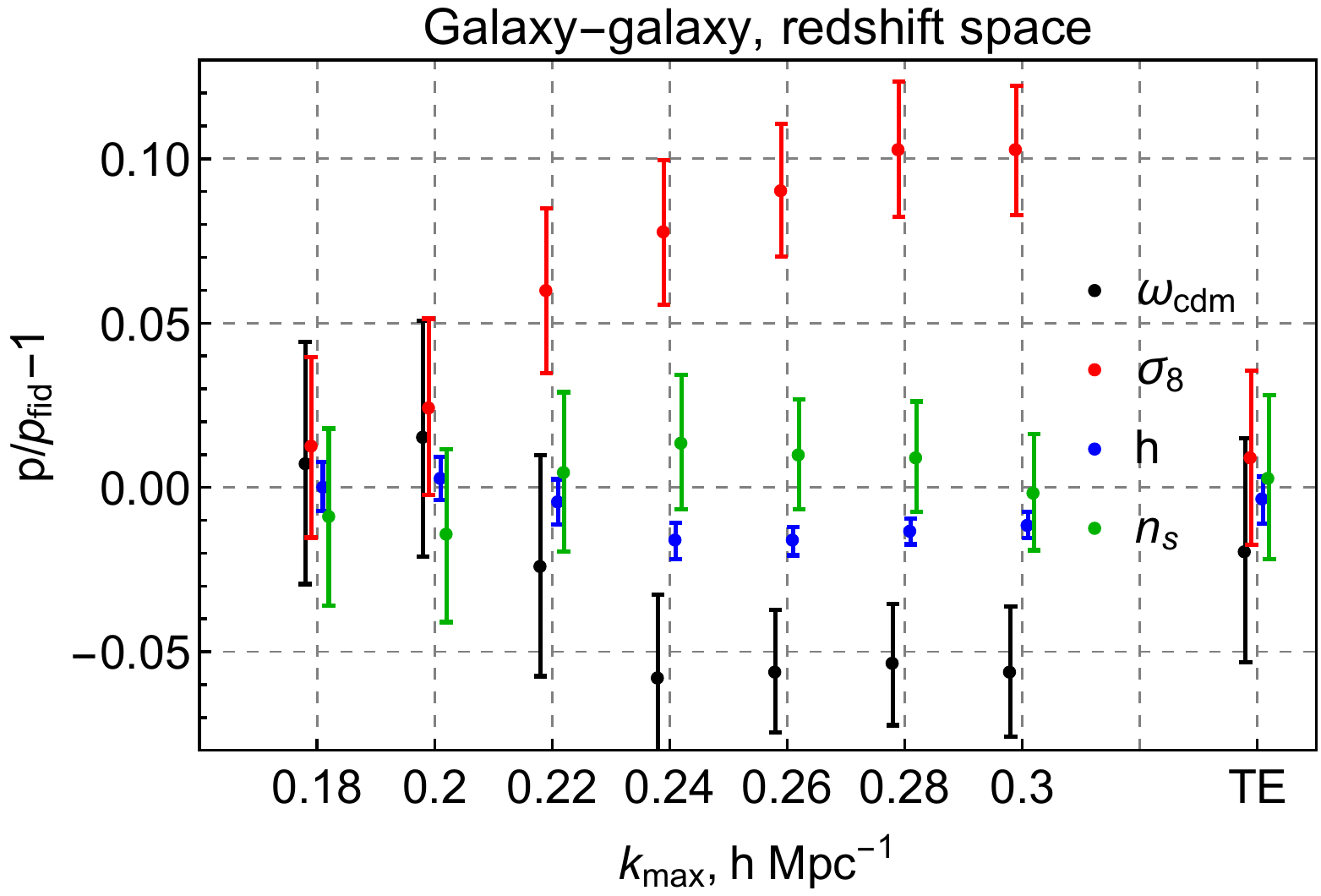}
	\caption[]{{\it Left panel:} Error budget of various systematic effects relative to the two-loop contribution (black line) at $z=0$. {\it Right panel:} Residuals in measurements of cosmological parameters from the redshift space galaxy multipoles of N-body data.}
	\label{fig:2}
\end{figure}

Finally, we discuss the systematic uncertainties associated with the implementation of IR resummation in the \texttt{CLASS-PT}. Detailed information about the implementation of IR resummation in \texttt{CLASS-PT} can be found in \cite{Chudaykin}. 
In the left panel of Fig. \ref{fig:2} we show the contributions of all these effect to the total error budget along with the two-loop correction at $z=0$.
Once can see that the errors caused by inaccuracies in IR resummation are always smaller than the two-loop contributions missed in our model. It means that the \texttt{CLASS-PT} provides a stable calculation of one-loop power spectrum up to next-to-leading order corrections. These corrections can be systematically 
taken into account within time-sliced
perturbation theory. 

\section{Theoretical error approach}
\label{sec:theor}

In this section we summarize the theoretical error approach~\cite{Chudaykin2,Baldauf:2016sjb}. 
We show how the theoretical error covariance can be included in the realistic data analysis.
\begin{enumerate}
	\item Choose some fiducial cosmological model.
	\item Select the fiducial data cut $k^{\rm fid}_{\rm max}$. This data cut should be reasonably small to make theoretical error negligible.
	\item Obtain the best-fit theoretical prediction $P^{\rm best-fit}(k)$ by fitting the data at $k^{\rm fid}_{\rm max}$ and by varying {\it only} nuisance parameters. 
    \item Take this best-fit theoretical curve and construct the theoretical envelope as 
    ${P}^{({\rm TE})}_i=P^{\rm d}_i - P^{\rm best-fit}(k_i)$.
    The statistical scatter in the data vector $P^{\rm d}_i$ can be removed by fitting ${P}^{({\rm TE})}_i$ with a smooth polynomial. 
    \item Build up the TE likelihood using ${P}^{({\rm TE})}_i$
    \be 
    -2\ln\mathcal{L}(P(\vec\theta))
    =(C+C^{\rm (E)})^{-1}_{ij}
    (P(\vec\theta)+\bar{P}^{({\rm TE})}_i-P^{\rm d}_i)
    (P(\vec\theta)+\bar{P}^{({\rm TE})}_i-P^{\rm d}_j)\,,
    \ee 
    where $\vec\theta$ denotes the vector of cosmological parameters and
    \be 
    C^{\rm (E)}_{ij}=
    \bar{P}^{({\rm TE})}_i
    \bar{P}^{({\rm TE})}_j e^{-\frac{(k_i - k_j)^2}{2\Delta k^2}}\,,
    \quad \text{with}\quad
    \Delta k=0.1~h/\text{Mpc}.
    \ee  
\end{enumerate}
We demonstrate the application of this approach using a suite of LasDamas Oriana simulations. Further
details can be found in \cite{Chudaykin2}.



We run several analyses with different $k_{\rm max}$ and compare these result with the TE approach. The marginalized constraints on cosmological parameters for the $k_{\rm max}$ and TE (rightmost point) analyses are shown in the right panel of Fig. \ref{fig:2}. We found that 
the TE analysis yields unbiased cosmological constraints with 
errorbars that match those 
coming from a carefully chosen 
$k_{\rm max}$. Thus, the TE appraoch effectively optimizes the choice of 
the data cut.

\section{Conclusion}
\label{sec:sum}

In this work we presented a new module \texttt{CLASS-PT} that incorporates one-loop theory calculations. It contains all ingredients needed for direct application to real data. We also introduced a new approach based on the theoretical error covariance which allows one to avoid uncertainties in the theoretical estimates of higher-order nonlinearities. We showed that this approach yields unbiased estimates of cosmological parameters and effectively optimize the choice of $k_{\rm max}$.  

\textit{Acknowledgments.}~We
are grateful to Mikhail Ivanov,
Oliver Philcox, and
Marko Simonovi\'c
for collaboration
on the projects presented in this note.




\begin{thebibliography}{99}


\bibitem{Chudaykin:2019ock}
A.~Chudaykin and M.~M.~Ivanov,
JCAP \textbf{11} (2019), 034,
arXiv:1907.06666 [astro-ph.CO]

\bibitem{Baumann:2010tm}
D.~Baumann, \textit{et al.}
JCAP \textbf{07} (2012), 051,
arXiv:1004.2488 [astro-ph.CO].


\bibitem{Blas:2015qsi}
D.~Blas, M.~Garny, M.~M.~Ivanov and S.~Sibiryakov,
JCAP \textbf{07} (2016), 052, arXiv:1512.05807 [astro-ph.CO].

\bibitem{Chudaykin}
A.~Chudaykin, M.~M.~Ivanov, O.~H.~E.~Philcox and M.~Simonovi\'c,
\Journal{\PRD}{102}{6}{2020},
arXiv:2004.10607 [astro-ph.CO].





\bibitem{Ivanov:2019pdj}
M.~M.~Ivanov, M.~Simonovi\'c and M.~Zaldarriaga,
JCAP \textbf{05} (2020), 042,
arXiv:1909.05277 [astro-ph.CO].

\bibitem{Simonovic:2017mhp}
M.~Simonovi\'c, T.~Baldauf, M.~Zaldarriaga, J.~J.~Carrasco and J.~A.~Kollmeier,
JCAP \textbf{04} (2018), 030, arXiv:1708.08130 [astro-ph.CO].


\bibitem{Blas:2016sfa}
D.~Blas, M.~Garny, M.~M.~Ivanov and S.~Sibiryakov,
JCAP \textbf{07} (2016), 028,
arXiv:1605.02149 [astro-ph.CO].

\bibitem{Ivanov:2018gjr}
M.~M.~Ivanov and S.~Sibiryakov,
JCAP \textbf{07} (2018), 053,
arXiv:1804.05080 [astro-ph.CO].
	
\bibitem{Chudaykin2}
A.~Chudaykin, M.~M.~Ivanov and M.~Simonovi\'c,
\Journal{\PRD}{103}{4}{2021},
arXiv:2009.10724 [astro-ph.CO].
	
\bibitem{Baldauf:2016sjb}
T.~Baldauf, M.~Mirbabayi, M.~Simonovi\'c and M.~Zaldarriaga, arXiv:1602.00674 [astro-ph.CO].



	




\end{thebibliography}
\end{document}